# Stiffening graphene by controlled defect creation

**Authors:** Guillermo López-Polín[1], Cristina Gómez-Navarro[1,2*], Vincenzo Parente[3], Francisco Guinea[3], Mikhail I. Katsnelson[4], Francesc Pérez-Murano[5], and Julio Gómez-Herrero[1,2]

[1] Departamento de Física de la Materia Condensada, Universidad Autónoma de Madrid, 28049, Madrid, Spain.

[2] Centro de Investigación de Física de la Materia Condensada, Universidad Autónoma de Madrid, 28049, Madrid.

[3] Instituto de Ciencia de Materiales, CSIC, 28049, Madrid, Spain.

[4] Radboud University Nijmegen, Institute for Molecules and Materials, Heyendaalseweg 135, NL-6525AJ Nijmegen, The Netherlands.

[5] Instituto de Microelectrónica de Barcelona, CSIC, 08193 Bellaterra, Spain.

*Correspondence to: cristina.gomez@uam.es

**Graphene extraordinary strength, stiffness[1] and lightness have generated great expectations towards its application in flexible electronics and as mechanical reinforcement agent. However, the presence of lattice defects, unavoidable in sheets obtained by scalable routes, might degrade its mechanical properties[2,3]. Here we report a systematic study on the elastic modulus and strength of graphene with controlled density of defects. Counter intuitively, the in-plane Young's modulus increases with increasing defect density up to almost twice the initial value for vacancy content of ~0.2%, turning it into the stiffest material ever reported. For higher density of vacancies, elastic modulus decreases with defect inclusion. The initial increase in Young's modulus is explained in terms of a dependence of the elastic coefficients with the momentum of flexural modes predicted for *2D* membranes[4,5]. In contrast, the fracture strength decreases with defect density according to standard fracture continuum models. These quantitative structure-property relationships, measured in atmospheric conditions, are of fundamental and technological relevance and provide guidance for applications in which graphene mechanics represents a disruptive improvement.**

Modifying the strength and stiffness of (3D) conventional materials by defect inclusion is a well-established technique in mechanical engineering. Reducing the dimensionality of the material usually entails a magnification of the influence of defects. In addition, sometimes reduced dimensionality brings up new emergent phenomena that have also to be considered. Graphene, a true 2D crystalline membrane of covalently bonded carbon atoms, has been shown to exhibit extraordinary intrinsic in-plane strength and Young's modulus[1], close to the elastic constant of the carbon covalent bonds. Experimental findings reveal a strong dependence of mechanical properties with defect content. A paradigmatic example is graphene derived from chemical reduction of graphene oxide that, due to its partial amorphous character, exhibits elastic modulus 5 times smaller than that of pristine graphene[2]. For polycrystalline graphene produced by chemical vapor deposition the effect of sample processing details in the grain boundaries significantly alter the elastic constants and strength of the sheets[3, 6, 7]. Unfortunately, the fact that these defects contents are not controlled but imposed by synthesis procedures and growth dynamics hinders systematic studies. A comprehensive approach to the role of disorder in graphene requires the introduction of defects in a controlled manner.

Therefore, in order to establish reliable structure-properties relationship the natural strategy is to begin with a pristine graphene sheet obtained by micro-exfoliation of natural graphite, for subsequent introduction of a known quantity of defects. Vacancies of carbon atoms are the simplest and more studied type of defect in graphene. Recently several theoretical works, performed using different approaches, [8-10] have predicted a decrease of both the 2D elastic modulus ($E_{2D}$) and strength with the introduction of such type of defects, according to what intuition dictates. As we shall demonstrate these calculations and naïve expectations fail. We show that, as a consequence of its out of plane fluctuations,[11] graphene $E_{2D}$ can be significantly increased by the inclusion of low density of defects in its atomic structure.

For this study graphene drumheads were prepared by mechanical exfoliation of natural graphite on Si(300nm)/SiO$_2$ substrates with predefined circular wells with diameters ranging from 0.5 to 3 µm (see SI1). The mechanical properties of the membranes were subsequently tested by indenting with an AFM tip at the center of the suspended area (details about the AFM probe can be found in SI2). AFM indentation experiments on graphene drumheads (figure 1) can be modeled as clamped circular membranes with central point load (see S4). The force vs. indentation curves can be approximated by the Schwering type solution as (equation 1) [1, 12].

$$F(\delta) = \pi\sigma_0\delta + \frac{E_{2D}}{a^2}\delta^3 \qquad (1),$$

where $F$ is the loading force, $\delta$ is the indentation at the central point, $\sigma_0$ is the pretension accumulated in the sheet during the preparation procedure and $a$ is the drumhead radius (see method section).

Results obtained in up to 30 pristine drumheads yielded values of $E_{2D}$ of 250-360 N/m (see SI4) and $\sigma_0$ ranging from 0.05-0.8N/m with no correlation between pre-stress and $E_{2D}$ (see Fig S8 in SI).

The fracture strength of the membranes was measured by loading some drumheads up to the failure point. The measured breaking forces ($F_{max}$) were ~1.7-2.1µN. Considering that the maximum stress under the tip can be expressed as $\sigma=(F_{max}E_{2D}/4\pi R_{tip})^{1/2}$, we obtain values of the breaking strength between 28 and 35 N/m. Summarizing, our elastic and strength values for pristine graphene are compatible to those reported previously in literature[1, 13, 14] ( see SI11).

With the aim of introducing a controlled density of defects the samples were irradiated with a known dose of Ar$^+$ with Energy of 140 eV (see methods and SI5). In these conditions a random spatial distribution of both mono- and di-vacancies of carbon atoms in a 3/1 ratio is created.[15, 16] The nature and density of defects were tested by Raman spectroscopy and scanning tunneling microscopy (STM). Figure 2a shows a representative image (acquired in atmospheric conditions) of a graphite sample irradiated in the same conditions as graphene flakes (see S7 for further details). The threefold perturbation observed at the center of the image identifies it unambiguously as an atomic point defect; Characterization by Raman spectroscopy reveal the in plane (sp$^2$) character of the defects. These two experimental findings together point toward either pure carbon vacancies or, most likely, vacancies chemically saturated by small atoms. STM imaging of the surface for periods of several days did not show any trace of image degradation by adsorption of airborne molecules.

Mechanical testing was performed after each irradiation dose with the same AFM probe and in the same conditions described above. Irradiated samples showed a similar $F$ vs. $\delta$ dependence to the non-irradiated ones. Panels b c and d of figure 2 gather representative data for a pristine membrane (as deposited) and a drumhead irradiated with a dose of *4x10$^{12}$ defects/cm$^2$*, corresponding to a mean defect distance of 5 nm. The free-of-defects membrane shows $E_{2D}$ of *305 N/m*, while the defected membrane shows a higher $E_{2D}$ (*484 N/m*). Figure 2d characterizes the measurements from a statistical viewpoint: the distance between Gaussians maxima is about 10 times their width.

In view of the above exposed results we proceeded to irradiate the samples in smaller steps. Colored points in figure 3a depict the results obtained for 3 representative drumheads that were characterized after each irradiation dose. Our main experimental finding is that the $E_{2D}$ of the graphene membrane increases with increasing irradiation dose and reaches a maximum of 550 N/m at a mean distance between defects of ~5 nm (0.2% defect content). For higher defect content we observe a decreasing $E_{2D}$. Figure 3a presents low data dispersion with a well-marked and very robust tendency. The pretensions in irradiated membranes undergo a slight increase but they are always below 0.8 N/m (figure SI8) that following previous reports[1] and detailed studies in our group (to be published elsewhere) do not justify by themselves the observed variations of $E_{2D}$.

With the aim of understanding the transition between pristine graphene and graphene oxide, in a recent report Zandiatashbar *et al.*[14] studied the dependence of graphene Young modulus subjected to oxygen plasma. While at first sight their results look contradictory with those reported here, detailed comparison points toward compatibility of the two experimental observations (see S16 for further details). As oxygen plasma simultaneously produces several type of defects, direct comparison of both works can only be made at very low defect densities, where the data are not contradictory. The region of defect densities where we observe the maximum in $E_{2D}$ is indeed not sampled in ref.[14] (S16). In addition, the appearance of large multi-vacancies in samples subjected to short periods of oxygen plasma justifies a pronounce drop in elastic modulus that should counteract our observed increase of the $E_{2D}$.

Although there is not a complete agreement in the theoretical predictions,[17, 18] most studies, performed either by molecular dynamics or density functional theory, do not predict elastic modulus enhancement with such a diluted density of defects,[8, 9, 19] but loss of rigidity (see SI12). None of these works considered the influence of defects in graphene within the framework of the thermodynamic theory of crystalline membranes[5, 11, 20] (SI9). Graphene exhibits a very low bending rigidity $\kappa$ of ~1eV comparable to that measured in biological membranes, where it is well known that entropic effects renormalizes the elastic constants. The low $\kappa$ introduces significant temperature ($T$) fluctuations and strong anharmonic effects that are important for wavelengths such as $q << q^* \prec \sqrt{(K_B T E_{2D})/\kappa^2}$ with a corresponding Ginzburg length[23] ($\lambda^* = 2\pi/q^*$) at $T=300$ K in the order of few nanometers. Anharmonicity is reflected in a strong coupling of in-plane and out-of-plane fluctuations giving rise to an exotic elasticity of the membrane, including the absence of any finite elastic constants in the thermodynamic limit, a negative Poisson's ratio, and thermal fluctuations characterized by a large anomalous dimension and negative thermal expansion. In particular, anharmonicity also introduces wavevector-dependent elastic modulus, $E_{2D} \propto q^{\eta_u}$ where $\eta$~0.36 and $q$ is the momentum or inverse length [4]. This dependence can be understood as a partial screening of the elastic coefficients due to the contribution from out of plane fluctuations to the free energy of the membrane (see SI9). Long wavelength excitations are favored in large and clean samples, where flexural modes have a long mean free path. According to our estimates (see SI9, SI10) defects lower the mean free path of flexural phonons and, eventually lead to their localization when $2\pi q^{-1}$~$d$, where $d$ is the mean distance between vacancies. As illustrated in fig. 4, the presence of defects suppresses flexural modes with longer wavelengths that do not contribute to the decrease of $E_{2D}$, leading to effective increase (see SI14, SI15). Note that we cannot discard a related scenario involving quenched ripples in suspended samples as observed in TEM images under electron irradiation.[21] Further support to our interpretation stem from our experimental observation that pristine membranes display enhanced $E_{2D}$ with increasing induced pre-stress (to be published) as predicted by the suppression of anharmonic effects in stiff membranes.[22]

This growth in the elastic coefficients will compete with the predicted softening effect of vacancies[8, 10, 19] and will be suppressed as the vacancy concentration approaches the percolation threshold where the elastic coefficients are expected to decrease linearly with the number of vacancies and will be proportional to the initial value of the coefficients. Taking into account these two competing mechanisms we can write a qualitative expression as:

$$E = K \cdot \left( b + \frac{1}{l_0^2} + n_i \right)^{\eta/2} \left( 1 - c \cdot \left( \frac{1}{l_0^2} + n_i \right) \right) \qquad (2)$$

where $K$ and $c$ are constants, $b$ is a geometrical factor of order of the inverse of the area of the drumhead that accounts for boundary conditions, $l_0$ is the localization length for flexural phonons in pristine graphene and $n_i$ is the density of defects induced by irradiation. The dashed line in figure 3a depicts a fitting to our experimental results according to equation 2. Best fitting to our experimental data yield $K=1.5 \times 10^9$ $Nm^{\eta-2}$, $l_0=70nm$ $\eta=0.36$ and $c=1.2 \times 10^{-18} \cdot m^2$. Since $1/l_0^2$ is much greater than $b$ (~$1/a^2$ being $a$ the hole radius) it can be neglected in the equation. From the fit we can conclude that the initial $E_{2D}$ in our samples is mainly dictated by the effective scattering length $l_0$~ 10-100 nm. This value is smaller than that reported for the localization length of flexural modes[23] at temperatures above 100 ºC; we attribute this difference to physisorption on our samples that desorbs at temperatures above 70ºC. This low value of $l_0$ does not allow experimentally observing the predicted dependence on the drumhead area. Our fitting value for the constant $c$ is higher than that predicted by molecular dynamics[10,19], this discrepancy can be attributed to defect agglomeration that might take place only at our higher irradiation doses as reported in ref[24]. In addition, vacancies lead to resonances at the Dirac energy and to the formation of a vacancy band with a high density of states; formation of such states result in a drop of elastic modulus.[25]

Interestingly, the average distance between defects at maximum $E_{2D}$ is about 5 nm (fig. 3a) that coincides fairly well with the above mentioned Ginzburg length beyond which flexural modes cease to be anharmonic. Hence, modes with shorter wavelengths do not screen the elastic constants.

Our findings imply that the applicability of atomistic models to mesoscopic membranes [26] might not be straightforward (see SI12 ans SI13 for a detailed discussion). In our experiments the variation of the Young's modulus takes place at relatively large vacancy distances (20-4 nm). Hence well averaged calculations should involve large slabs with more than 10000 atoms. Although these sizes can be reached with molecular dynamic simulations, the result of these simulations will always depend on the accuracy of the selected potential. Precise DFT calculations beyond perturbation theory, fully including thermal fluctuations of flexural phonons, electron-phonon coupling and atom slabs as large as possible, should address our observations.

Note also that our proposal together with the experimental data implies non renormalized $E_{2D}$ greater than *550 N/m*. Experimental data on pre-strained graphene membranes (to be published) yield even higher values *i.e.~700 N/m*. The lower $E_{2D}$ measured by nanoindentation in irradiated samples *vs.* these obtained in pressurized membranes[22] suggests that defects do not totally cancel the entropic effects.

Finally, measurements of the breaking force showed a marked different behavior when compared to $E_{2D}$. The fracture force of irradiated membranes displays a pronounced decrease upon defect creation. Data are depicted in Figure 3b. The measured failure forces drop in a factor of two for the lower irradiation dose, corresponding to a mean distance between defects of 12 nm. This corresponds to a 30% reduction of strength (see SI11). For higher defect density the value of the breaking force presents a high dispersion varying from 400 to 800 nN. We do not observe an additional significant drop even for irradiation doses as high as $4 \times 10^{13}$ defects/cm$^2$ (~1%), indicating a saturation tendency. In contrast to the behavior of the $E_{2D}$ with defect density, which cannot be explained by conventional continuum mechanics, the fracture strength dependence can be addressed by classical models.[9]

Our determination of the extreme relevance of defects on the mechanical properties of suspended graphene sheets brings up fundamental issues in 2D materials that can be extrapolated to other atomic thin membranes, such as BN and MoS$_2$. The provided elasticity and strength *vs.* defect concentration relationships, measured in ambient conditions, open new paths to tailoring the stiffness of future graphene based devices. The introduction of such amounts of defects should enhance the sensitivity of mass sensors based on graphene resonators and should be taken into account for using graphene as reinforcement agent.

**Methods.**

Figure 1 illustrates sample geometry. Single layers of graphite were found by optical microscopy and corroborated by Raman spectroscopy [27] (see SI). Non contact AFM images of the drumheads showed that graphene layers adhere to the vertical walls of the wells for 2-10 nm in depth (see Figure 1d). The experiments described in this work were carried out in ambient conditions. Only membranes showing a flat and featureless surface (i.e. absence of bubbles or wrinkles) and not noticeable slack were selected for the measurements. Repeated loading/unloading cycles on the same membrane (see SI3) showed high reproducibility, completely reversible behavior and no signature of fatigue.

Obtaining $E_{2D}$ from equation 1 requires a precise determination of the tip-membrane contact point that is not always easy to determine from experimental curves. In order to check the robustness and validity of indentation curves we have used 3 different equations to fit the indentation curves (see S4 for detailed information). Differences between these fittings are within 20% in absolute value but always yield a similar tendency. For the values reported here $E_{2D}$ was estimated from the coefficient in $\delta^3$ of a complete third order polynomial (mode 3 in described in S4).

The density of defects was estimated by two independent methods: measuring the ionic current density as a function of time and performing Raman spectroscopy after each irradiation dose. The mean distance between defects was deduced from the intensity relation between the *D* and *G* peaks in Raman spectra.[24, 28] Both methods yield similar results. Raman spectroscopy[29] and scanning tunneling microscopy (STM) characterization at atomic level (see Figure 2a, SI6 and SI7) also corroborated the vacancy-like character of the induced defects, ruling out out-of-plane chemisorption in air. AFM images of the drumheads and surrounding regions after irradiation showed no differences compared to images prior to irradiation. We have not observed signatures of plasticity (i.e. irreversible effects) up to the forces reached in our indentations.

Our results are robust respect to tip variations. We measured with different probes finding always the same tendency and similar values as those shown in fig. 3a. Consecutive indentations with the same tip in pristine-irradiated-pristine samples yielded consistent $E_{2D}$ enhancement in irradiated samples. The analysis described in SI1 is accurate for indentation curves with maximum forces above 200 nN. This threshold is much smaller than the maximum force used for pristine (800 nN) and irradiated (600 nN) samples.

**Acknowledgments:** This work was supported by. MAT2013-46753-C2-2-P, Consolider CSD2010-0024, FIS2011-23713, and the European Research Council Advanced Grant, #290846. We acknowledge technical support from A. Aranda, C. Salgado, A. del Campo, and fruitful discussions with M. Jaafar, A. K. Geim, R. Perez, F. Yndurain and J. Soler.

**Author contributions**
C. G-N and J.G-H devised the experiments. G.L-P performed the experiments. G.L-P, C. G-N and J.G-H analyzed the data. F.P-M prepared the substrates. C. G-N and J.G-H wrote the manuscript. V.P., M.I.K. and F.G. formulated the theoretical model. All author participated in discussions.

**Figure legends**

**Fig. 1. Set up and sample geometry.** (**a**, **b**) Device geometry and scheme of the nanoindentation and irradiation set-up. (**c**) Optical microscopy image of a mono-(1), bi-(2) and multi-layer (N) graphene contacted by a gold electrode deposited on an array of wells. Scale bar: 15µm. (**d**) AFM image (top view) of a graphene sheet covering several circular wells. The graph corresponds to the topographic profile along the green line in the image.

**Fig. 2. Characterization of pristine and defective graphene.** (**a**) STM image acquired in air at room temperature of a defect created on a graphite sample irradiated in the exact same conditions described in the main text. The image is in excellent agreement with those reported previously for single atomic vacancies[17]. The electronic perturbation near the defects observed as a threefold periodicity surrounding the defect identifies it unambiguously as a point defect. Color code applies to panels b c and d: blue represents measurements on a graphene membrane as deposited. Red corresponds to measurements on the same drumhead as blue data after irradiation with $4·10^{12}$ defects/cm$^2$. (**b**) Raman spectra of the same membrane before and after irradiation. Red curve is shifted in the vertical axis for clarity. (**c**) F(δ) obtained in the same drumhead before (blue) and after (red) irradiation. (**d**) Histogram of the $E_{2D}$ obtained from several indentations in the same drumheads as previous panels. In order to compare with three dimensional materials, in the upper axis we divide by the interatomic layer spacing in graphite (0.34 nm).

**Fig. 3. Mechanical characterization as a function of defect density.** (**a**) $E_{2D}$ as a function of defect concentration. Green circles represent two representative drumheads (drumhead 1 and drumhead 2) that were tracked after multiple irradiations. Grey squares represent measurements of up to 20 different drumheads contained in 5 different membranes. Blue and red circled stars correspond to the $E_{2D}$ obtained from the measurements depicted in figure 2. Error bars are not shown for the sake of clarity. Measurements with errors bars can be found in SI4.   Dashed line is a fit to equation 2 in the main text. (**b**) Breaking Force for multiple

drumheads as a function of created defect density. Error bars correspond to standard deviation of measurements performed on different drumheads. The green dashed line is just a guide to the eye.

**Fig. 4. Thermal fluctuation in 2D membranes. (a)** Fluctuations of a suspended graphene membrane in the absence of strain. (**b**). In the presence of defects, long wavelength fluctuations are quenched

FIGURE 1

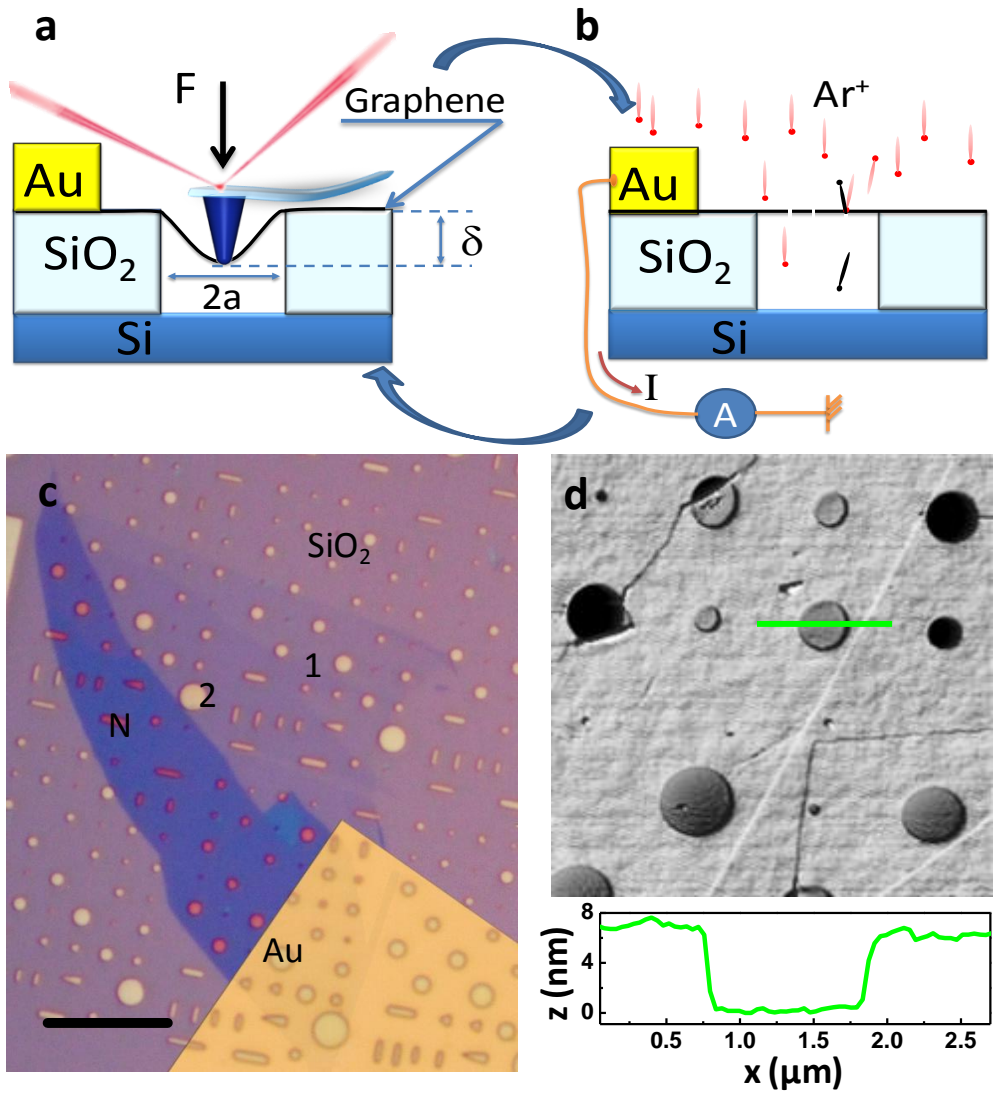

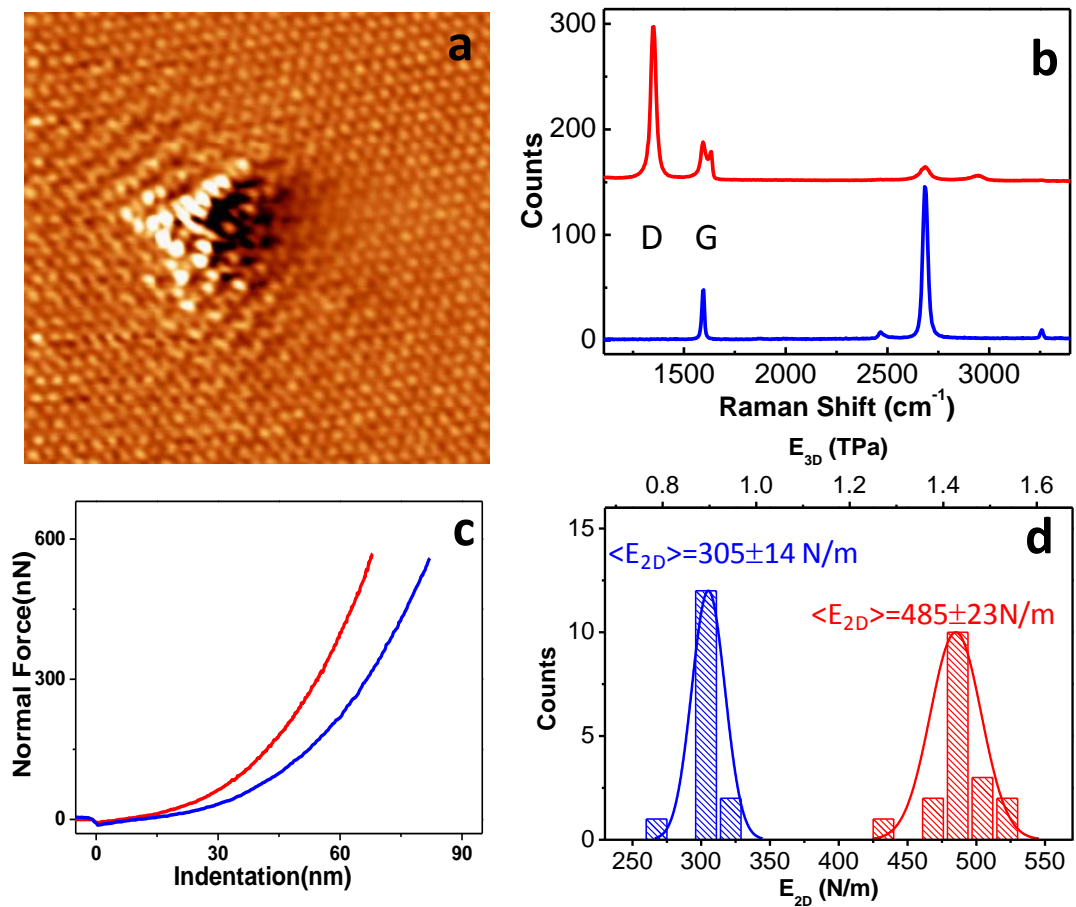

FIGURE 3

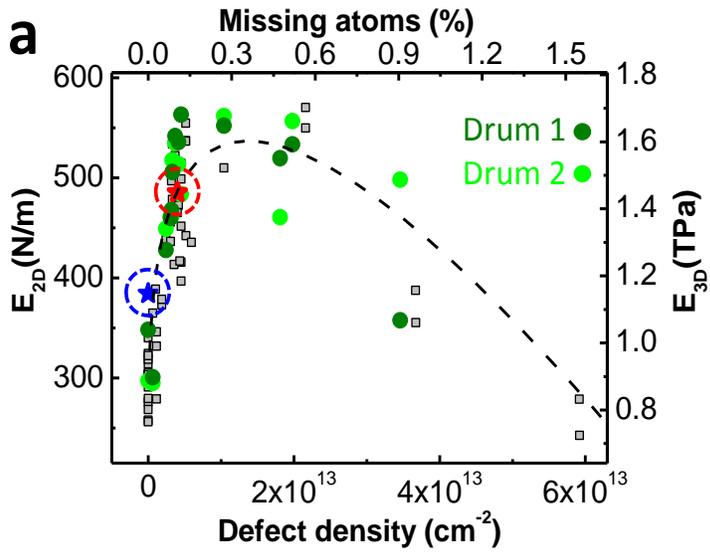

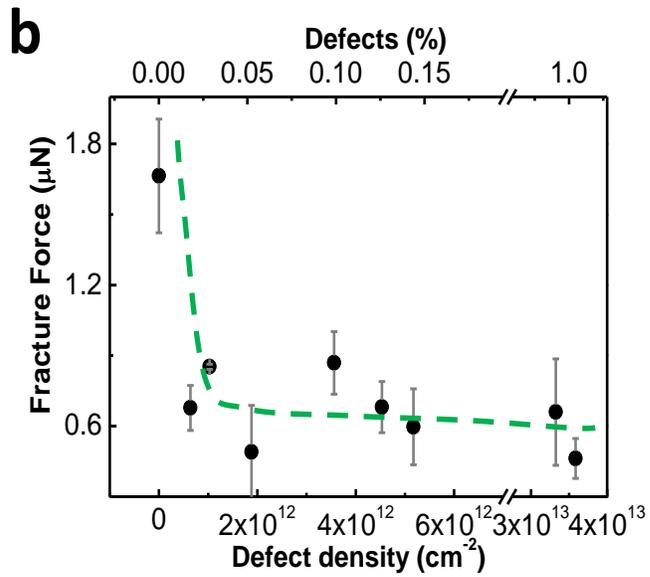

**FIGURE 4**

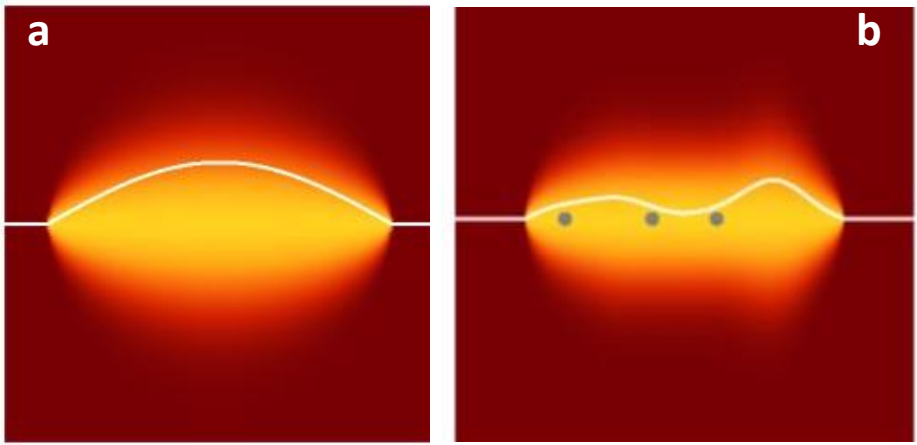

**Short accessible summary for homepage**

Graphene is one of the stiffest materials ever reported exhibiting a Young's modulus comparable to that of diamond.  In contrast, its bending rigidity is very low, similar to that of soft lipid layers. This unique combination of properties makes it highly interesting.  Low density of defects induced by ion irradiation has now been observed to enhance graphene Young's modulus up to twice its pristine value. This increase can be understood in the framework of the thermodynamic theory of crystalline membranes where the elastic coefficients are predicted to depend on the momentum of flexural modes.

# Stiffening graphene by controlled defect creation

## Supplementary Information

**Authors:** Guillermo López-Polín, Cristina Gómez-Navarro*, Vincenzo Parente, Francisco Guinea, Mikhail I. Katsnelson, Francesc Pérez-Murano, and Julio Gómez-Herrero

*Correspondence to: cristina.gomez@uam.es

**Index.**
S1 . Substrate fabrication
S2: AFM tip characteristics.
S3: Force vs indentation curves.
S4: Determination of $E_{2D}$ from experimental data.
S5: $Ar^+$ irradiation conditions.
S6. Raman spectroscopy.
S7: Scanning Tunneling Microscopy (STM) of graphite before and after Ar irradiation.
S8: Pretension as a function of irradiation dose.
S9 Qualitative description of the effect of flexural modes on the in-plane elastic constants.
S10. Lifetime of flexural modes and localization threshold.
S11. Strength determination and comparison with previously reported data.
S12. Detailed comparison with existing calculations.
S13. Relation between membrane theory and DFT calculations. Acoustic phonon lifetimes.
S14. Perturbation analysis of the Young's and bulk modulii.
S15. Renormalization Group Approach.
S16 $Ar^+$ irradiation vs. oxygen plasma.

**S1: Substrate fabrication.**

The substrates were fabricated from 6 inch silicon wafers (As-doped, resistivity 1 – 3.5 mΩ·cm). A 300 nm thick silicon oxide was grown by dry oxidation. The back of the wafer was coated with a 0.8 µm thick layer of AlSi(1%)Cu(0,5%).

The silicon oxide was patterned using projection optical lithography and reactive ion etching (RIE). A chip was designed with patterns of different shapes (fig. S1) and with critical dimensions of 0.5 µm, 1 µm, 1.5 µm and 3 µm, and separations between patterns of 2.5 µm and 3.5 µm. The design pattern was repeated many times in the chip, together with additional navigation marks, in order to easily locate the position of the individual graphene drumheads by optical microscopy and by AFM. The dimensions of the chip are 4 mm x 4 mm.

For the optical lithography process, the wafers were coated with a 0.6 µm thick photoresist and exposed using an i-line stepper. After development of the resist, the wafers were cleaved in pieces containing 4 x4 chips. The RIE process was performed in an AMS system using a mixture of gases ($C_4F_8$: 30 sccm and CH₄: 20 sccm ) for 45 s, in order to fully remove the silicon oxide in the areas without photoresist. The silicon oxide etch rate for this recipe was measured to be 400 nm/min. After the RIE, the resist was removed by oxygen plasma (power and time). Then, the pieces were subjected to an additional cleaning step by immersing in a stripper solution (Microstrip® 2001, Fuji.film) at 60 ºC with ultrasounds for 10 minutes, rinsed in DI water for 10 minutes, and finally submitted to another oxygen plasma cleaning process to remove any remaining residual on the surface.

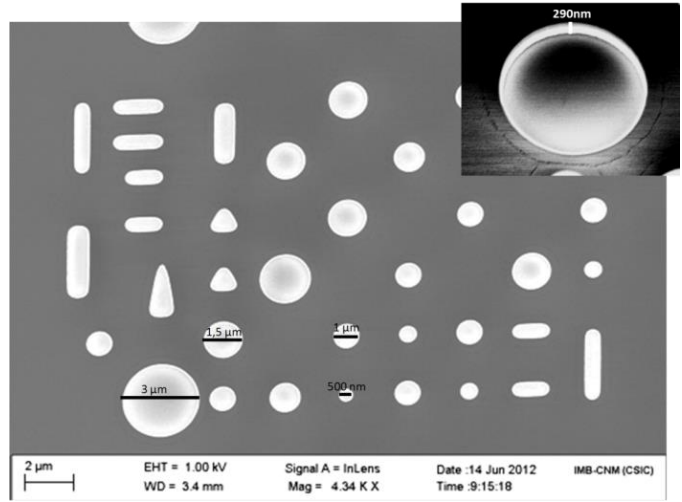

**Figure S1** shows a SEM image of one part of the chip after all the processes and a zoom in one of the holes to appreciate the verticality of the walls.

## S2: AFM tip characteristics.

In order to have a constant and well defined contact geometry we use commercial tips from NanoScience Instruments with hemispherical geometry and low wear coating of Tungsten carbide with nominal final tip radius of 60 nm (fig. S2). Previous attempts to obtain reproducible results with standard Si or slender diamond tips failed because the high stiffness of graphene membranes produced tip breaking. Cantilever spring constant is 33 N/m calibrated by Sader method[1].

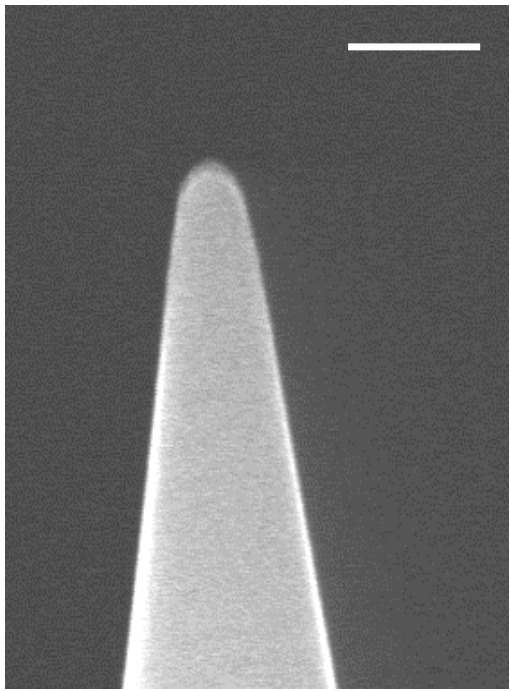

**Figure S2 (**Scale bar 200 nm). SEM image of the AFM tip used for the measurements.

We checked that the increase in $E_{2D}$ occurs for different type of cantilevers with force constants ranging from 2 up to 40 N/m and tips made of silicon, diamond and tungsten carbide. In other words, the in plane stiffening is independent of the probe used for the experiments.

## S3: Force vs indentation curves.

Indentation experiments were performed using a Nanotec AFM with the WSxM software[2] package. Although our results did not depend on the approaching/retracting speed in the range of 1-1000 nm/s, all the measurements presented here were obtained at the same loading/unloading rate of 90nm/s. In order to use equation 1 in the main text, indentation has to be accurately estimated. Indentation is not a direct experimental measure; it is calculated from the differences of the relative displacement of the samples and the tip on the non-deforming $SiO_2$ substrate. This is illustrated in figure S3.

In order to calculate indentation it is then critical to fix the zero displacement point, or zero force level: Inaccuracy of 2-5 nm in this point leads to a 10% error in the final calculated $E_{2D}$. To solve this problem we have developed an iterative numerical method that sweeps the contact point position. For each position $E_{2D}$ is calculated using conventional least square fitting. We take as the optimum $E_{2D}$ the one that minimizes the fitting error.

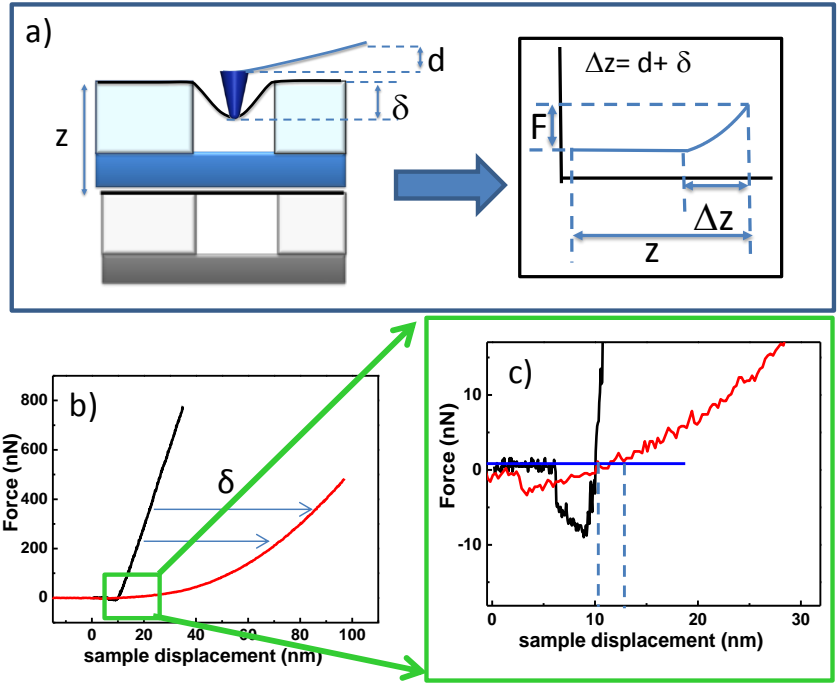

**Figure S3**.a) scheme illustrating the experimental procedure to extract the indentation (δ) of the membrane from the displacements of the cantilever (d) and a reference substrate (Δz). b) Force *vs.* sample displacement curve acquired on the SiO$_2$ substrate (black) and on the center of a suspended sheet (red). Panel c depicts a zoom of the same curve in the green rectangle. Here we can appreciate that, due to the flat character of the curves near the zero force level, we have an experimental accuracy of 2-5 nm in determining the position of zero deflection.

## S4: Determination of E$_{2D}$ from experimental data.

We fitted the experimental force vs. indentation curves using 3 different criteria:

1. We find a *first order* zero for the curve as the point where the horizontal dashed line cut the curve (zero cantilever deflection). A zoom in this region shows the experimental noise of the curve. To avoid noise as much as possible we find the zero by a purely instrumental fitting of the data to a high order polynomial (3$^{th}$ order). To this end we take about 20 points to the left and about 70 to right of the initial zero. E$_{2D}$ is then estimated according to equation 1 in the main text

2. We find a *first order* zero and then we scan around this number. For each zero we fit a polynomial with a linear and a cubic coefficient and we calculate $S=\Sigma(F_i-C_1\delta-C_3\delta^3)^2$. We use to calculate the Young's modulus the $C_3$ from the polynomial that minimize $S$.

3. We obtain the Young's modulus from the $C_3$ coefficient of a full third order polynomial. In this case the position of the zero is irrelevant.

For the three methods described above we always obtain the same tendency for the Young's modulus as a function defect density. The precise $E_{2D}$ for each method presents variations within a 20%. However the most robust value is the one obtained according to method 3, therefore we have selected this methodology for our estimations. Our study are in agreement with that reported in ref[3].

While our dispersion in results for a given pristine membrane is quite low (about 10 N/m, (see Figure 2c in main text), we do find bigger differences in $E_{2D}$ when comparing different drumheads and our estimated $E_{2D}$ for 20 different drumheads vary from 270 to 370 N/m. We attribute this scattering to several sources.

1-We partially attribute the variability between different drumheads to the experimental differences in the boundary conditions. Eq 1 assumes a circular membrane perfectly clamped along its perimeter. The holed structures present in our substrates were performed by optical lithography. Therefore, we expect nanometer sized features that differ from a perfect circle and would slightly vary actual boundary conditions. We should also consider that clamping to the substrate requires a minimum contact area between the graphene sheet and the $SiO_2$ surface. Graphene adheres to $SiO_2$ by van der Waals forces. According to the values reported in[4] we have estimated the minimum area of a graphene ring required to sustain a force of 1µN (maximum force applies in our indentations). This yields a ring width of approximately 5-10 nm, introducing another uncertainty in the clamping boundary conditions.

2-Another source of error in $E_{2D}$ can be an inexact determination of hole diameter with AFM images due to tip convolution. Notice that the dependence in $\delta^3$ is proportional to $E_{2D}/a^2$. Therefore a 5% error in the determination of *a* translates into a 10% in the estimated $E_{2D}$.

3-Finite size of indenter: as stated in the main text equation 1 describes the $F(\delta)$ behavior for a point load on a finite membrane. A complete and detailed analysis of the validity of this equation for different indenter sized is describe on ref[5]. Here it is shown that the point load equation yields best results for a $R_{tip}/a \cong 0.05$. For higher (lower) values, the $E_{2D}$ estimated from this fits is over (infra) estimated. However we don't observe any dependence of the estimated $E_{2D}$ with well diameter, therefore we conclude that the previously described sources of dispersion are more likely to account for our observations.

Remarkably, none of these sources of dispersion can justify the observed increment in $E_{2D}$ at low irradiation density.

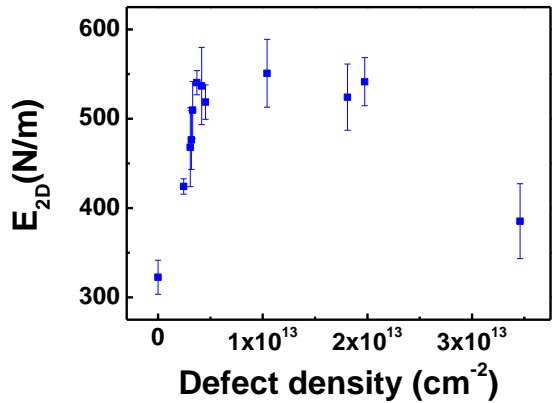

**Figure S4.** $E_{2D}$ obtained for multiple F(δ) curves for a representative drumhead that was measured after each irradiation dose. The error bars correspond to the standard deviation obtained for multiple F(δ) curves.

### S5: Ar$^+$ irradiation conditions.

Irradiation of the samples (fig S5) was performed in a HV chamber with a base pressure of 1x10$^{-7}$ mbar. The Ar$^+$ pressure during irradiation was 5x10$^{-5}$ mbar. Ar$^+$ energy was 140 eV. The distance between the sputtering gun and sample was 30 cm. The sample was electrically contacted to an aluminum circular plate of 3.5 cm of diameter. The sputtering spot was large enough to fully cover the supporting plate. In order to measure the ionic current and to fix an electrostatic reference voltage, the graphene flakes were previously contacted with a microscopic gold electrode. Measuring the current between sample and ground yields the rate of ionic bombardment that was around 3x10$^{-8}$A/cm$^2$. Irradiation time was between 10 s and 3min. Assuming that each argon ion removes a carbon atom (both calculations[6-8] and precise STM measurements[9] indicate that this is the case) it is possible to estimate the density of defects produced by the irradiation for a given period of time. Since the ionic current varies slightly with time, in our experiments we use a data acquisition board to measure the ionic current as a function of time. The numerical integral of the acquired data yields the total charge dose hitting the surface and hence the total number of carbon atoms etched from the surface.

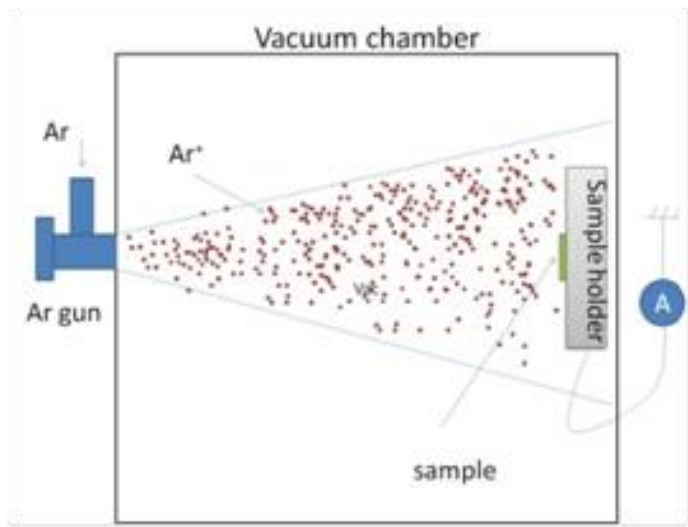

**Figure S5** : Argon irradiation set-up and geometry

### S6: Raman spectroscopy.

Raman spectra were performed using a WITEC/ALPHA 300AR Raman confocal microscope at ambient conditions. The laser wavelength and power were 532nm and 0.7mW respectively.

### Determination of the number of layers.

Graphene monolayers were first identified by optical microscopy and then corroborated by Raman spectroscopy as described in ref [10](fig. S6.1)

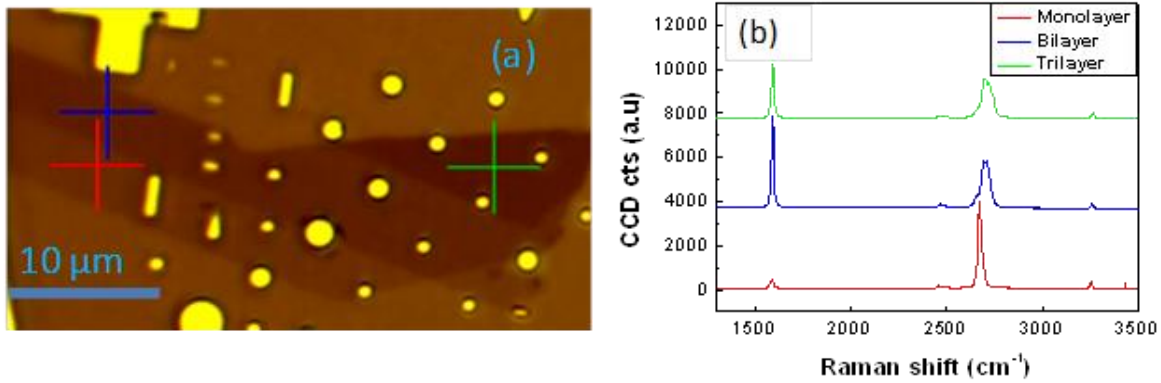

**Figure S6.1.** Example of a graphene structure displaying mono- bi- and tri- layer regions.

### Determination of defect density.

Maps of Raman spectra of the entire graphene flakes were acquired to ensure the homogeneity of the flake. In order to estimate the $I_D/I_G$ relation we acquire average spectrum always on the same region of the flake.

The mean distance between defects after each irradiation dose ($L_D$) is estimated according to the expression given in [11]:

$$\frac{I_D}{I_G} = C_A \frac{(r_A^2 - r_S^2)}{(r_A^2 - 2r_S^2)} \left( e^{-\pi r_S^2/L_D} - e^{-\pi(r_A^2 - r_S^2)/L_D} \right)$$

Where we consider $r_S$=1nm, $r_A$=3.1 nm and $C_A = AE_L^{-4}$, where $E_L$ is the laser energy and A=180 eV$^4$.

### Determination of the nature of defects

The vacancy-like nature of defects was determined by the relation between the D and D' peaks of the Raman spectra as stated by Eckman et al.[12].

- $I_D/I_{D'} \approx 13$ -> sp$^3$ defects
- $I_D/I_{D'} \approx 7$ -> sp$^2$ defects
- $I_D/I_{D'} \approx 4$ -> boundary defects

In figure S6.2 we plot the $I_D/I_G$ vs $I_{D'}/I_G$. From the slope of the linear fit we extracted $I_D/I_{D'}$. The value so obtained was ~6,5. Implying the presence of clean vacancies, instead of sp3-type defects (oxidized, hydrogenated, fluorinated...). In the case of $sp^3$ type defects this relation should be around 13. This analysis allows us discarding the presence of chemisorbed molecules on the graphene surface.

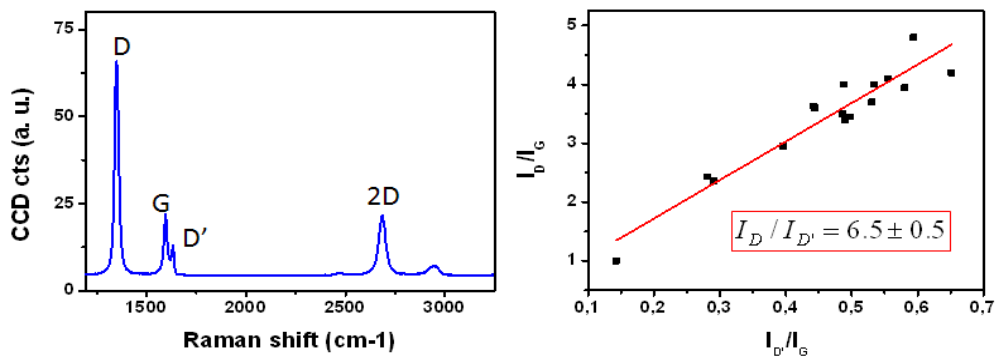

**FIGURE S6.2** a) Raman spectrum where D and D' peaks can be observed. b) Values obtained for $I_D/I_G$ vs $I_{D'}/I_G$.

**S7: Scanning Tunneling Microscopy (STM) of graphite before and after Ar irradiation.**

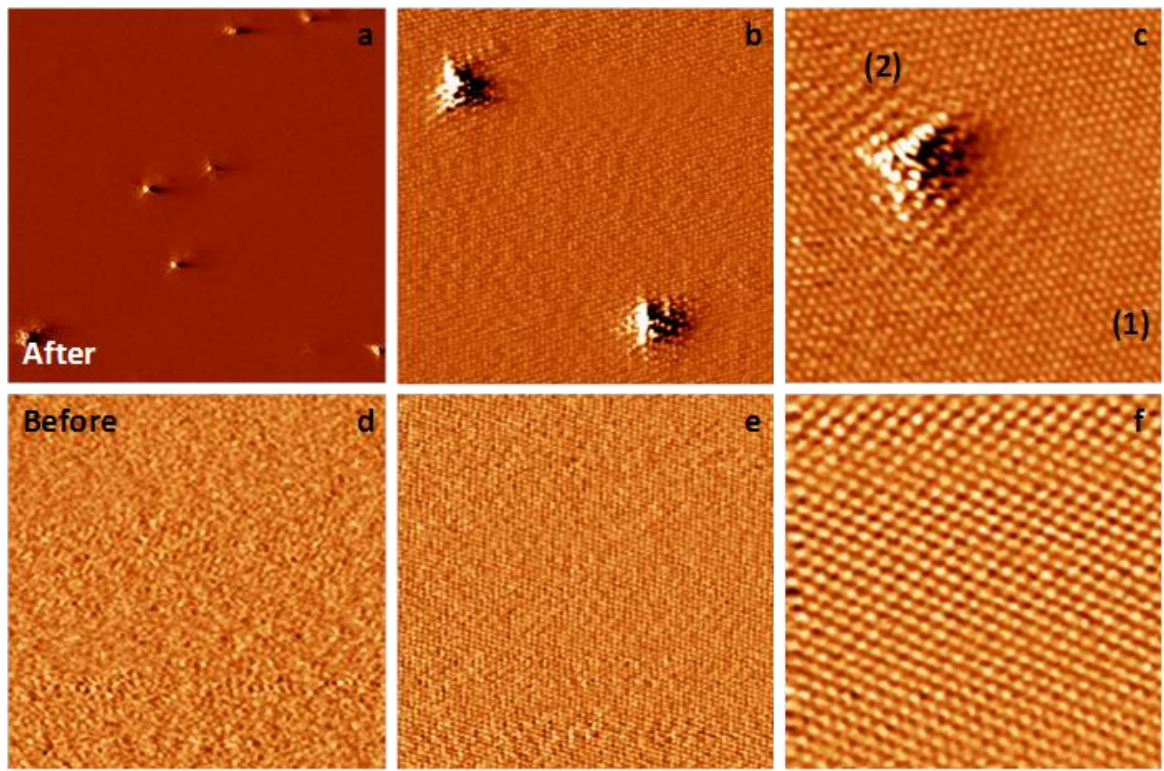

**Figure S7.** *Atmospheric STM images of graphite before (lower images) and after (upper images) irradiation with Ar ions with the same conditions used for graphene. Image sizes are: panels a and d 35x35nm2, panels b and e*

*15x15nm², panel c 8.5x8.5nm2and panel f 6.3x6.3nm². Region (2) in panel c corresponds to the threefold $\sqrt{3}x\sqrt{3}$ perturbation due to the defect, in contrast to the hexagonal atomic periodicity typically observed by STM in pristine graphite (region 1) of the same panel.*

In order to further corroborate the punctual vacancy nature of defects created by Argon irradiation in the conditions described in section S5, we irradiated HOPG graphite samples in these conditions and imaged them before and after irradiation by STM in air-ambient conditions.

Fig. S7 displays STM images of graphite before and after irradiation in the exact same conditions used in the experiments described in the manuscript. While images in graphite prior to irradiation show perfect atomic lattice, atomically resolved images on irradiated samples reveal small defects. These defects, visualized as small protrusions by STM, are uniformly distributed all through the sample. On the regions between defects we always observe a clear and perfect atomic periodicity corresponding to that typically observed by STM in pristine graphite. Our high resolution images of defects (fig. S7c) are in excellent agreement with those reported previously for single atomic vacancies[9,13]. Furthermore, the electronic perturbation near the defects observed as a threefold periodicity surrounding defects identifies them unambiguously as point defects (i.e. smaller than lattice spacing). The irradiate graphite surface was scanned in air ambient conditions by STM during 3 consecutive days and we did not observe any trace of image degradation by airborne molecules.

## S8: Pretension as a function of irradiation dose.

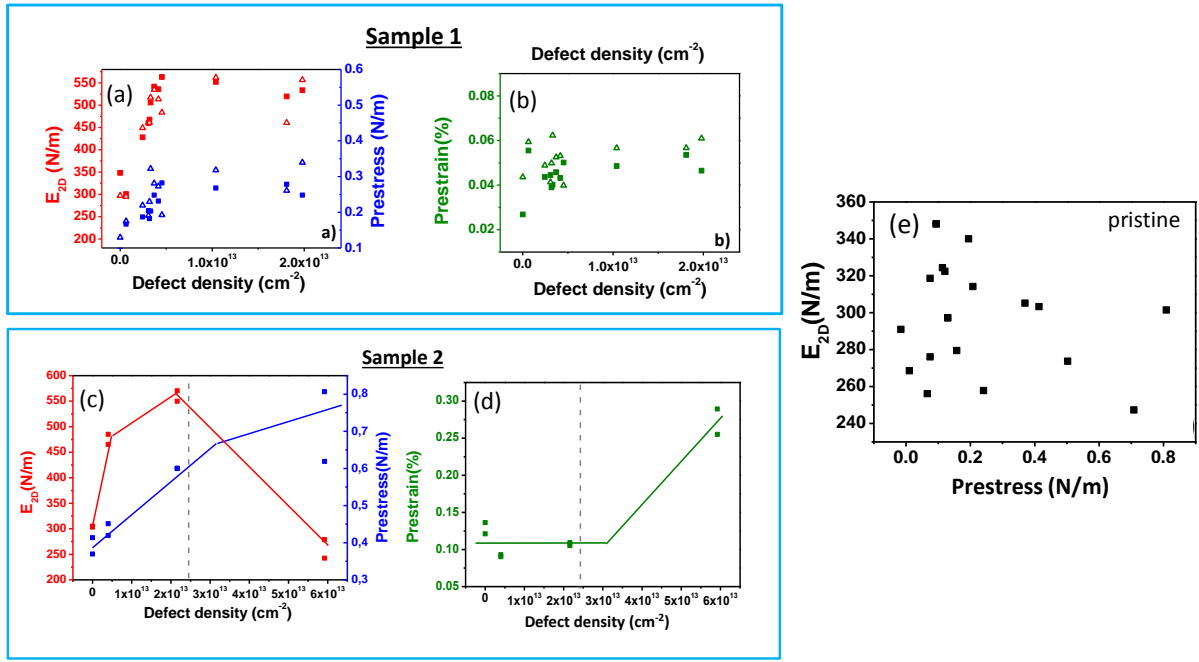

**Figure S8** *a) Values obtained for $\sigma_0$ and $E_{2D}$ for two representative drumheads as a function of introduced defect density. B) Prestrain for the same two drumheads as in a). c) and d) Same as a) and b) but for a sample with higher irradiation. The dashed line indicates the maximum irradiation reached in sample 1. Continuous lines are drawn to guide the eye. For sample 2 the prestress keeps growing while the stiffness*

*starts to drop. e) Elastic modulus as a function of pre-tension for pristine (non-irradiated samples) showing no correlation.*

The pre-strain present in the membrane prior to indentation experiments is calculated by fitting our F(δ) curves to a complete third order polynomial. For the pre-tensions in irradiated sample we observe a tendency similar to that of $E_{2D}$. Remarkably, if we calculate the strain in %, as $\varepsilon=\sigma_0/E_{2D}$ we find that it is constant with the irradiation dose (fig S8).

More significant is the fact that in pristine sheets we do not observe any correlation between the calculated $\sigma_0$ and $E_{2D}$, meaning that the origin of this effect is not in the analysis, but intrinsic to the samples (fig. S8e).

**S9: Qualitative description of the effect of flexural modes on the in-plane elastic constants.**

We give herein a simple explanation of the renormalization of the Young's modulus by thermally excited flexural modes shown in fig. 4 of the main manuscript. We consider a membrane of linear dimension $l$. The full expression for the elastic energy is

$$E = \frac{\lambda}{2}\int d^2\vec{r}\left(\partial_x u_x + \frac{(\partial_x h)^2}{2} + \partial_y u_y + \frac{(\partial_y h)^2}{2}\right)^2$$

$$+ \mu \int d^2\vec{r}\left[\left(\partial_x u_x + \frac{(\partial_x h)^2}{2}\right)^2 + \left(\partial_y u_y + \frac{(\partial_y h)^2}{2}\right)^2\right.$$

$$\left.+ \frac{1}{2}\left(\partial_y u_x + \partial_x u_y + (\partial_x h)(\partial_y h)\right)^2\right] + \frac{\kappa}{2}\int d^2\vec{r}(\partial_x^2 h + \partial_y^2 h)^2$$

Where $\lambda$ and $\mu$ are the elastic Lamé coefficients, $\kappa$ is the bending rigidity, the vector $(u_x, u_y)$ gives the in plane displacements, and $h$ is the out of plane displacement. We consider fluctuations in $h$ with wavelengths comparable to $l$. We study the perturbative regime, where the fluctuations are dominated by the bending rigidity, $\kappa \gg \lambda l^2, \mu l^2$. In this limit, the amplitude of the flexural modes is determined only by the bending rigidity. Their contribution to the partition function is

$$Z_0 \propto \int dh\, e^{-c_0\kappa(h^2/l^2)/T} \propto \sqrt{\frac{T}{c_0\kappa/l^2}}$$

Where $c_0$ is a numerical constant of order unity. The free energy is

$$F_0 = -T\log Z_0 = -T\log\sqrt{\frac{T}{c_0\kappa/l^2}}$$

An applied strain $\bar{u}$ couples to the flexural modes through the Young's modulus, $E_{2D}$. The free energy becomes

$$Z \propto \int dh\, e^{-(c_0\kappa(h^2/l^2)+c_1 E_{2D}\bar{u}h^2)/T} \propto \sqrt{\frac{T}{c_0\kappa/l^2 + c_1 E_{2D}\bar{u}}}$$

Where $c_1$ is another numerical constant. The free energy is now

$$F = -T\log Z = -T\log\sqrt{\frac{T}{c_0\kappa/l^2\,(1 + c_1 E_{2D}\bar{u}l^2/c_0\kappa)}}$$

The Young's modulus acquires a correction given by

$$\delta E_{2D} \propto \frac{1}{l^2}\frac{\partial^2 F}{\partial \bar{u}^2} \sim -\frac{T E_{2D}^2 l^2}{\kappa^2}$$

By replacing the dimension by the inverse momentum, $l \leftrightarrow q^{-1}$, this simple estimate gives a parametrically correct expression for the renormalization of the Young's modulus in the weak coupling regime, $[TE_{2D}(q)]/(q^2\kappa^2) \ll 1$, ([14])

$$\frac{q}{E_{2D}(q)}\frac{\partial E_{2D}(q)}{\partial q} \propto \frac{T E_{2D}(q)}{q^2 \kappa^2(q)}$$

A related scaling equation can be obtained for the bending rigidity, $\kappa(q)$

$$\frac{q}{\kappa(q)}\frac{\partial \kappa(q)}{\partial q} \propto -\frac{T E_{2D}(q)}{q^2 \kappa^2(q)}$$

When the parameters of the material do not satisfy the weak coupling condition, as in graphene, the scaling equations outlined here need to be replaced by the Self consistent Screening Approximation, SCSA. This analysis leads to the scaling laws described in the main text. Note that the bending rigidity, $\kappa$, increases at low wavevector, while the Young's modulus, $E_{2D}(q)$, decreases. Hence, for sufficiently low wavevectors, $q \ll q_c \approx \sqrt{(TE_{2D})/\kappa^2}$ the scaling equations outlined above are valid. If we assume that $E_{2D}(q) \propto q^{\eta_u}$ and $\kappa(q) \propto q^\eta$ and insert these expressions in the scaling equations above, we obtain the consistency condition $\eta_u = 2 - 2\eta$, in agreement with other approaches [14,15].

**S10: Lifetime of flexural modes and localization threshold.**

We consider first the point defects with infinite mass. At the position of these defects, we have $h = 0$. We can take into account this constraint by including in the free energy a term

$$\delta E = U \sum_i \int d^2\vec{r}\, h(\vec{r})^2 \delta(\vec{r} - \vec{r}_i)$$

Where the coordinates $\{\vec{r}_i\}$ define the positions of the defects, and $U \to \infty$ is a scale with dimensions of energy x area. For a single defect, the Green's function satisfies

$$G(\vec{q},\vec{q}',\omega) = G_0(\vec{q},\omega) + \sum_{\vec{q},\vec{q}''} G_0(\vec{q},\omega) U G(\vec{q}'',\vec{q}',\omega)$$

Where

$$G_0(\vec{q}, \omega) = \frac{1}{\rho\omega^2 - \kappa|\vec{q}|^4}$$

We obtain

$$\sum_{\vec{q}} G(\vec{q}, \vec{q}', \omega) = \frac{\sum_q G_0(\vec{q}, \omega)}{1 - UA^{-1}\sum_q G_0(\vec{q}, \omega)}$$

Where $A$ is the area of the system. We finally obtain

$$G(\vec{q}, \vec{q}', \omega) = G_0(\vec{q}, \omega) + \frac{\sum_{\vec{q},\vec{q}'} G_0(\vec{q}, \omega) U G_0(\vec{q}', \omega)}{1 - UA^{-1}\sum_q G_0(\vec{q}, \omega)}$$

From this equation we can define a self energy for a system with a concentration $c$ of defects

$$G(\vec{q}, \vec{q}', \omega)^{-1} = \left(G_0(\vec{q}, \omega) - \Sigma(\vec{q}, \omega)\right)^{-1}$$

Where

$$\Sigma(\vec{q}, \omega) = \frac{c}{\sum_q G_0(\vec{q}, \omega)}$$

And

$$\sum_q G_0(\vec{q}, \omega) \approx \frac{1}{\kappa|\vec{q}|^2}$$

So that $\Sigma(\vec{q}, \omega) \approx c\kappa|\vec{q}|^2$. The threshold for localization is given by $\kappa|\vec{q}|^4 \approx \Sigma(\vec{q}, \omega)$, that is, $q_{loc} \sim \sqrt{c}$. Flexural modes are localized, and do not contribute to the renormalization of the Young's modulus, when their wavelength is comparable to the distance between defects.

The boundary condition for vacancies is such that tension at their boundary should vanish. This condition implies that $|\partial h| = 0$ at the position of the vacancy. We can satisfy this boundary condition by adding a term to the energy

$$\delta E = V \sum_i \int d^2\vec{r}\left[(\partial_x h)^2 + (\partial_y h)^2\right]\delta(\vec{r} - \vec{r}_i)$$

Where $V$ is a large scale with dimensions of energy. We apply the same analysis as for the case of defects of infinite mass and obtain

$$\Sigma(\vec{q}, \omega) = \frac{c|\vec{q}|^2}{\sum_q |\vec{q}|^2 G_0(\vec{q}, \omega)} \approx \frac{c\kappa|\vec{q}|^2}{\log(q_T/|\vec{q}|)}$$

Where $q_T \sim \sqrt{(TE_{2D})/\kappa^2}$ is a high momentum cutoff, and, as before, the self energy is independent of $V$. The threshold for localization is given by $q_{loc} \approx \sqrt{c}/\log(q_T/\sqrt{c})$. This cutoff is similar to the one obtained for infinite mass defects, except for a logarithmic correction.

**S11: Strength determination and comparison with previously reported data.**

Strength was estimated according to the simplified expression for a linear elastic model as $\sigma=(F_{max}E_{2D}/4\pi R_{tip})^{1/2}$ where $F_{max}$=1.7-2.1 µN, $R_{tip}$=50-60nm and $E_{2D}$=300-340N/m. These values yield $\sigma$=28-33 N/m. Previous works[16,17] reported values of 50-45 N/m where corrections due to non-linear effects lead to final values of 42-35 N/m.

The main difference between these reports and our work is that these experiments were carried out using single crystal diamond tips. This diamond tips tend to exhibit crystallographic planes (i.e. well faceted structures) where the spherical indenter model used in the above expression might not be so suitable , since the contact are between the tip and graphene layer (where the maximum stress is produced) is not so well characterized. Indeed according to our experimental experience performing fracture indentation experiments in graphene membranes, small changes in tip geometry yields changes much higher than those derived from non linear correction. In addition, another possible source of variability on the reported strength is the difference in chemical reactivity depending on tip material [18], that could also translate in different breaking stress for tungsten carbide and diamond. A more precise determination of the strength would require a deeper analysis and most likely a different experimental set up, where the breaking force is not influenced by tip details. Both of them are beyond the scope of this manuscript.

Therefore, in order to have consistent and robust comparative values for the breaking forces of pristine and defective graphene membranes, the data reported in fig 3b (main manuscript) were all acquired with the same AFM tip, in addition we first perform test (failure) experiments on pristine membranes and then, after breaking defective ones and observing an accused drop in the breaking force, we carry out again experiments in pristine graphene drumheads with the same tip. These *round trip* experiments exhibit variations on the breaking force of pristine graphene less 15%, indicating that the tip shape is not significantly altered by repeated indentations. Note that the main focus in this work is not the determination of an absolute value of the intrinsic strength but a comparative study between pristine and defective graphene.

**S12: Detailed comparison with existing calculations.**

As discussed in the main text, a number of calculations based on Density Functional Methods (DFT) report a Young's modulus for graphene consistent with the observed experimental value, $\approx 1$ TPa. Other DFT calculations overestimate, however, the Young's modulus of the closely related material fluorgraphene[19].

DFT calculations of the elastic properties of graphene have been recently extended to the study of the anharmonic properties[20,21]. In agreement with the analysis reported here, these calculations show a relevant coupling between in plane and out of plane modes at long wavelengths, leading to a finite lifetime for the in plane acoustic modes at infinite wavelength[19-21]. As the dispersion relation of these modes is determined by the elastic constants, the existence of a finite lifetime implies an imprecision in the calculation of the Young's modulus, when DFT is extended in order to include anharmonic effects.

A significant theoretical literature exists on the elastic properties of graphene with vacancies. Methods include the use of force models and DFT[22-32]. The vast majority of these calculations suggest that vacancies lead to lower values for the elastic constants. This is consistent with the assumption that the extraordinary stiffness of garphene is due to the robustness of the C-C σ bonds. To a first approximation, the effect of vacancies is to reduce the number of σ bonds in the structure. None of these calculations include [22-32] the effect of thermal fluctuations, which can be relevant at low vacancy concentrations, as discussed in this manuscript.

A very detailed analysis of a geometry similar to the one studied experimentally here is presented in[22]. Due to limitations imposed by the numerical method, the dimensions of the systems studied are $\approx 15$ nm, and the lowest vacancy concentrations are $\approx 1\% \approx 2 \times 10^{13}$ cm$^{-2}$. For this range of concentrations, a homogeneous decrease of the Young's modulus as function of vacancy concentration is found, in agreement with the experimental results presented here.

## S13: Relation between membrane theory and DFT calculations. Acoustic phonon lifetimes.

With the aim of further supporting the continuum theory used in this case, in the following, we discuss the relations between the continuum theory of membranes and recent calculations based on Density Functional Theory on the strength of anharmonic effects in graphene and graphite [19,20].

We analyze the coupling between in-plane acoustic phonons and out of plane flexural modes. We reproduce the main result in refs 19 and 20; the existence of a finite lifetime for the long wavelength in plane modes in lowest order perturbation theory. We give a simple analytical estimate of this lifetime in terms of the macroscopic elastic constants of the membrane. Finally, we show how the calculation can be extended to the study of the modification of the sound velocity induced by anharmonic effects. These results further support the claim made in the main text that the elastic properties of graphene at long wavelength cannot be rigorously defined even by DFT unless the details of the sample and the experimental setup are specified.

The coupling between in-plane modes and flexural modes can be expressed as:

$$H_{coupling} = \frac{\lambda}{2} \int d^2 \vec{r} (\partial_x u_x + \partial_y u_y)\left((\partial_x h)^2 + (\partial_y h)^2\right)$$
$$+ \mu \int d^2 \vec{r} \left[\partial_x u_x (\partial_x h)^2 + \partial_y u_y (\partial_y h)^2 + (\partial_y u_x + \partial_x u_y)(\partial_x h)(\partial_y h)\right]$$

We now quantize this expression using creation and destruction operators associated to the excitations of the harmonic system. The in plane modes can be divided into transverse and longitudinal phonons. For the transverse phonons we find, after some algebra:

$$H_T = \mu \sum_{\vec{k},\vec{q}} q^2 k \sin(2\phi) \frac{1}{\sqrt{2\sqrt{\rho\mu k}}} \frac{1}{2[\sqrt{\rho\kappa}(k^2 - q^2)]} (b_{\vec{k}}^{+T} + b_{-\vec{k}}^{T})\left(b_{(\vec{q}-\vec{k})/2}^{+F} + b_{-(\vec{q}-\vec{k})/2}^{F}\right)\left(b_{(-\vec{q}-\vec{k})/2}^{+F} + b_{(\vec{q}+\vec{k})/2}^{F}\right)$$

where $\rho$ is the 2D mass density, and $\phi$ is the angle between the vectors $\vec{k}$ and $\vec{q}$. Using second order perturbation theory, the inverse lifetime of the transverse phonons is

$$\Gamma_T = \frac{\mu^{3/2}}{16\pi\rho^{3/2}\kappa} \int q\,dq\,d\phi \frac{q^4 k \sin^2(2\phi)}{(k^2 - q^2)^2} \delta\left(\sqrt{\frac{\mu}{\rho}}k - \sqrt{\frac{\kappa}{\rho}\frac{(k^2+q^2)}{2}}\right)\left(\frac{8T(k^2+q^2)}{\sqrt{\frac{\kappa}{\rho}}(k^2-q^2)^2}\right)$$

where the delta function ensures that the energy of the transverse mode is equal to the sum of the energies of the two flexural modes in which it decays, and the last term stands for the thermal population of the flexural modes. As $k \to 0$ the delta function implies that $q^2 \to 2\sqrt{\mu/\kappa}\,k$. We finally obtain:

$$\Gamma_T = \frac{\mu T}{4\rho^{1/2}\kappa^{3/2}}$$

In a similar way, the coupling to the longitudinal modes is:

$$H_L = \sum_{\vec{k},\vec{q}} k[\lambda(k^2 - q^2) + 2\mu(k^2 - q^2\cos^2(\phi))] \frac{1}{\sqrt{2\sqrt{\rho\mu k}}} \frac{1}{2[\sqrt{\rho\kappa}(k^2-q^2)]} \left(b_{\vec{k}}^{+L} + b_{-\vec{k}}^{L}\right)\left(b_{(\vec{q}-\vec{k})/2}^{+F}\right.$$
$$\left. + b_{-(\vec{q}-\vec{k})/2}^{F}\right)\left(b_{(-\vec{q}-\vec{k})/2}^{+F} + b_{(\vec{q}+\vec{k})/2}^{F}\right)$$

and

$$\Gamma_L = \frac{1}{16\pi\rho^{3/2}\sqrt{\lambda + 2\mu}\,\kappa} \int q\,dq\,d\phi \frac{k[\lambda(k^2 - q^2) + 2\mu(k^2 - q^2\cos^2(\phi))]^2}{(k^2 - q^2)^2} \delta\left(\sqrt{\frac{\lambda + 2\mu}{\rho}}k\right.$$
$$\left. - \sqrt{\frac{\kappa}{\rho}\frac{(k^2+q^2)}{2}}\right)\left(\frac{8T(k^2+q^2)}{\sqrt{\frac{\kappa}{\rho}}(k^2-q^2)^2}\right)$$

Finally, we obtain, for $k \to 0$

$$\Gamma_L = \frac{(\lambda + \mu)^2 T}{4\rho^{1/2}(\lambda + 2\mu)\kappa^{3/2}}$$

### S14: Perturbation analysis of the Young's and bulk modulii.

Calculations in the previous section can be extended and the corrections to the real part of the transverse and longitudinal phonons can be estimated. One need to make the replacement in the integrals for $\Gamma_T$ and $\Gamma_L$:

$$\delta\left(v_{L,T}k - \sqrt{\frac{\kappa}{\rho}\frac{(k^2+q^2)}{2}}\right) \to \frac{1}{2\pi\left(v_{L,T}k - \sqrt{\frac{\kappa}{\rho}\frac{(k^2+q^2)}{2}}\right)}$$

As in any calculation in second order perturbation theory, the correction to the energies is negative. In the limit $k \to 0$ the $q$ integrals diverge as $q^{-2} \sim k^{-1}$. This divergence is another manifestation of the strong anharmonic effects in membranes in the long wavelength macroscopic limit.

For simplicity, we discuss only the correction to the real part of the self energy of the transverse acoustic modes. We obtain

$$\text{Re}\Gamma_T = \delta\left(\sqrt{\frac{\mu}{\rho}}k\right) = \frac{\delta\mu}{2\mu}\sqrt{\frac{\mu}{\rho}}k = -\frac{\mu^{3/2}Tk}{2\pi\rho^{1/2}\kappa^2}\frac{1}{q_c^2}$$

where $q_c$ is the inverse of a short wavelength cutoff due to the experimental set up. The change in $\mu$, $\delta\mu$ becomes

$$\frac{\delta\mu}{\mu} = -\frac{\mu T}{\pi\kappa^2 q_c^2}$$

in agreement with the general theory of membranes. This expression is valid if the term in the r.h.s. is much smaller than 1.

A more exact treatment requires the use of the Self Consistent Harmonic Approximation, SCHA. The displacement of a D dimensional membrane embedded in a d dimensional spaces is parameterized by a D dimensional vector describing the in-plane distortions and a $d_c$ =d-D dimensional vector describing the out-of-plane oscillations of the membrane. Via a $1/d_c$ expansion it is possible to derive a set of self-consistent equations whose numerical solution allows determining the height-height correlation function and the Young's modulus. Fig. S14 shows the increase in the bulk modulus, calculated within the SCHA, when a finite concentration of infinite mass defects is described by means of a self energy $\Sigma(\vec{q},\omega) \approx c\kappa|\vec{q}|^2$, where $c$ is the concentration of defects.

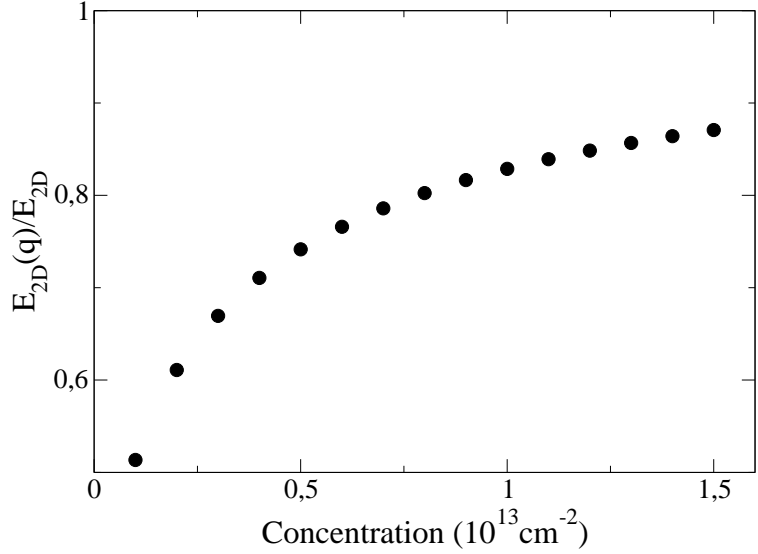

**Figure S14**. Bulk modulus divided by the zero temperature bulk modulus of a membrane as function of the concentration of infinite mass defects (see text).

**S15: Renormalization Group Approach.**

The calculation shown in fig. S14 can be compared with a Renormalization Group calculation (RG in the following). The integration of the in plane modes in the elastic action leaves the effective action for the flexural phonons

$$S = \frac{1}{2}\int \frac{d^2q}{(2\pi)^2}\frac{d\omega}{2\pi}(\rho\omega^2 - \kappa q^4)h(q,\omega)h(-q,-\omega) - \frac{1}{2}\int \frac{d^2q}{(2\pi)^2}\frac{d\omega}{2\pi}E_{2D}(q)u(q,\omega)u(-q,-\omega)$$

where $u(q,\omega) = (1/2)P_{ij}\partial_i h \partial_j h$, and $P_{ij}$ is the transverse projector, and the function $E_{2D}(q) = (E_{2D})_0 + (E_{2D})_1 q + (E_{2D})_2 q^2$ is the Young's modulus. The RG analysis can be performed to highlight the low energy behavior of the $K_n$ coefficients, through the introduction of the high energy cut-off $q_c$. The bare couplings $K_n$ are renormalized through a bubble like diagram involving two exural modes given by the diagram shown in fig. S15.1.

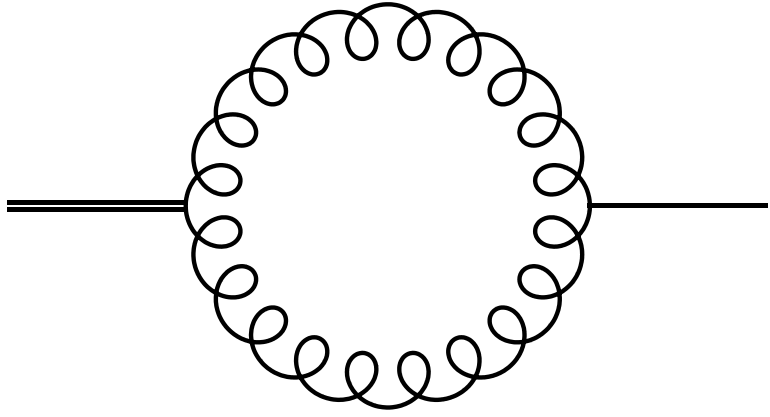

**Figure S15.1** Polarization diagram renormalizing the out of plane modes of a membrane.

This diagram is logarithmically divergent in the cut-off momentum $q_c$, so corrections can be re-adsorbed in the effective couplings leading to the scaling equations for q going to zero

$$q_c \frac{\partial (E_{2D})_n}{\partial q_c} = \frac{3\hbar}{64\pi} \sum_{i+j=n} \frac{(E_{2D})_i (E_{2D})_j}{\sqrt{\rho} \kappa^{\frac{3}{2}}}$$

$$q_c \frac{\partial \kappa}{\partial q_c} = -\frac{3\hbar}{16\pi} \sum_i q_c^i \frac{(E_{2D})_i}{\sqrt{\rho} \kappa^{\frac{3}{2}}}$$

The effect of vacancies is to actually change the cut off $q_c$: in pristine samples $q_c$ is determined through the inverse of the size of the sample, since it's the only length scale of the problem. When vacancies are produced another length scale is introduced, namely the average distance between the defects, and thus increasing the number of defects is equivalent to increasing the cut off $q_c$. As shown in fig. S15.2 the bulk modulus at zero temperature is enhanced for growing $q_c$ leading to a quick saturation to the pristine value $K_0$. The Equations regualting the renormalization of the Young's modulus with the infrared cut-off cannot be solved analytically in general but on the assumption $\kappa \approx \kappa_0$ the equation for $K_0$ can be integrated giving the result

$$(E_{2D})_0 (q_c) = \frac{(E_{2D})_0}{1 - K_0 A \log \frac{q_c}{q_D}}$$

where $q_D = 2\pi/a$ is the Debye momentum, and

$$A = \frac{3\hbar}{64\pi} \frac{1}{\sqrt{\rho} \kappa^{3/2}} = 1.21 \times 10^{-4} \dot{A}^2/eV$$

At finite temperature the equation describing the renormalization of the Young's modulus independent of momentum $K_0$ is

$$q_c \frac{\partial (E_{2D})_0}{\partial q_c} = \frac{3\hbar}{64\pi} \frac{(E_{2D})_0^2}{\sqrt{\rho} \kappa^{3/2}} \coth \frac{l_t^2 q_c^2}{2}$$

The solution of this equation is reported in fig. S15.2.

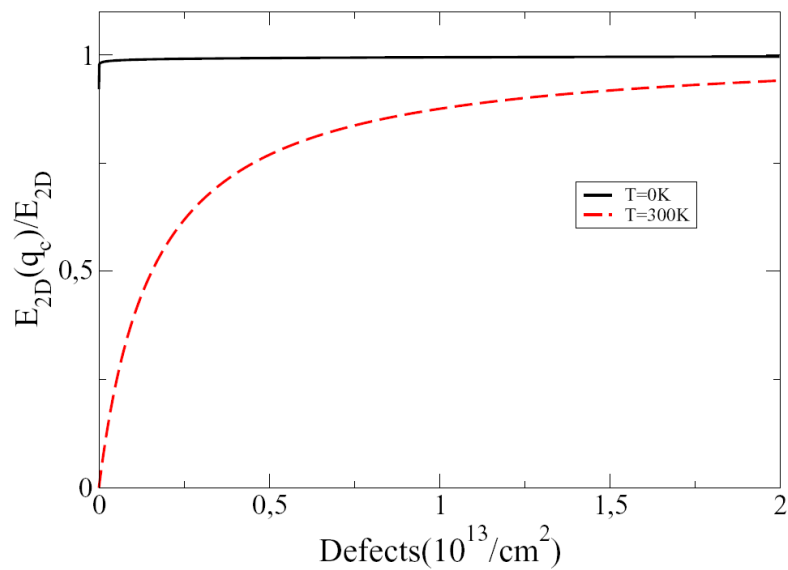

**Figure S15.2**. The renormalizad Young's modulus at 300K (dotted red line) compared with the zero temperature solution.

The increase in Young's Modulus found in the calculation shown in the last three sections are in good agreement with that found experiemtally. This calculations do not consider the softening trend.

**S16: Ar$^+$ irradiation vs. oxygen plasma.**

In the following lines we compare our results obtained using Ar ion irradiation with those reported by Zandiatashbar *et al.* [17].

According to ref. [17] oxygen plasma inherently produces several type of defects simultaneously (oxygen decorated atoms and multi-vacancies). This fact impedes a "clean" monitoring of any magnitude as a function of oxygen decoration. In addition, while Argon irradiation allows controlling the quantity of defects by in situ measurement of the current reaching the sample (see S5 supplementary information), oxygen plasma does not allow uniform sampling due to the lack of an analogous magnitude to be tracked.

In order to directly compare our results and those reported in ref. [17] we have used chart 4a combined with chart 2d (both in ref [17]). In Figure 4a they plot the Young's modulus and Raman relation peaks as a function of plasma time. According to our experience with Raman and to previous studies in literature (see ref. [12] and [11] summarized in S6 in SI) we have converted the horizontal axis in this graph to defects density and we have marked the nature of created defects inferred from their data. More in detail, figure 2d (ref [17]) shows the evolution of the defect type as a function defect average distance. From this plot it is evident that even at very low plasma time (~13 nm average distance between defects) the ratio I(D)/I(D') is 10 indicating that they have already a coexistence of sp$^3$ defects with sp$^2$ nanopores. Also from this plot we can conclude that for average defect distances shorter than 4-5 nm (20 s plasma time) the I(D)/I(D') ratio drops below 7, that is the threshold for sp$^2$ type defects. Hence, for longer time (shorter distances) they do not have sp$^2$ type defects but something that according to [12] is more related to boundary defects.

Our results and these reported by Zandiatashbar are plotted simultaneously in figure S16a of this document, figure S16c is a zoom on the interesting region. By inspecting this figure there are two clear issues that deserve consideration:

1) The creation of large multi-vacancies already takes place at very low oxygen plasma doses. This is inherent to the technique, making the coexistence of both type of defects (sp$^3$ oxygen sites and large multi-vacancies) unavoidable even at low densities. This fact prevents for systematic studies as a function of just one type of defect. More importantly, it is well accepted that the presence of large multi-vacancies results in a pronounced drop of the Young's modulus counteracting the effect that we report. In fact, at very low oxygen doses, where the multi-vacancies are not present, their data seem to show an increase in the Young's modulus. However the interval is so narrow that it might not be significant.
2) Data sampling is not uniform in the case of oxygen plasma. There is a wide gap of experimental points in the region where we observe the maximum of Young's modulus. This irregular sampling is also inherent to oxygen plasma since there is not any in situ magnitude to be tracked and defect creation with time is not uniform. Notice that Raman experiments for determination of defect density are *ex situ*. Indeed, there is a broad literature by Krasheninnikov and coworkers (see for instance ref.[6] and reference therein) modelling the results of Ar$^+$ irradiation in carbon nanotubes and graphene. There is nothing similar to that for oxygen plasma.

In summary, in our experiments with Ar irradiation we create a single type of defects (mono-vacancies); In the oxygen plasma case even at low doses several type of defects are created and for high doses the nature of defects becomes unpredictable. In addition, the so called vacancy type defects,[17] correspond to large holes that significantly decrease $E_{2D}$, hiding the effect that we report. Importantly, for intrinsic reasons the creation of defects by oxygen plasma etching is quite irregular in time, leaving a large void in the region where we observe the increase in $E_{2D}$.

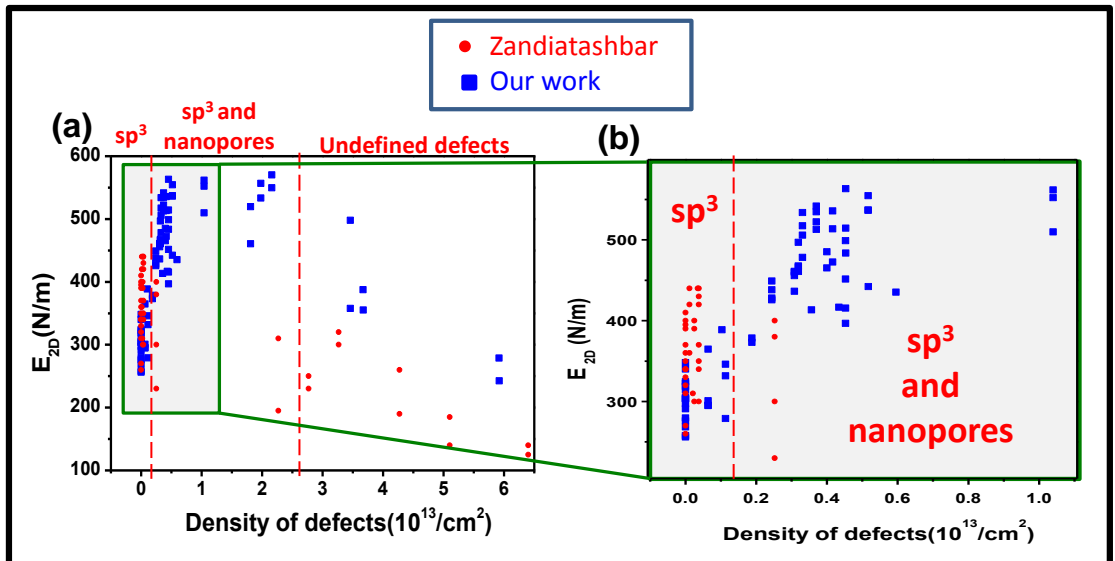

**Figure S16.** a) $E_{2D}$ vs. defects density for both (ref) experiments (red dots and labels) and our experiments (blue squares). b) zoom in the enclosed region